# AN APPROACH FOR STITCHING SATELLITE IMAGES IN A BIGDATA MAPREDUCE FRAMEWORK

H. Sarı [a], S. Eken [a], A. Sayar [a]

[a] Kocaeli University, Computer Engineering Department, Kocaeli, Turkey – hnisa_sari@hotmail.com,(suleyman.eken, ahmet.sayar)@kocaeli.edu.tr

**ABSTRACT:**

In this study we present a two-step map/reduce framework to stitch satellite mosaic images. The proposed system enable recognition and extraction of objects whose parts falling in separate satellite mosaic images. However this is a time and resource consuming process. The major aim of the study is improving the performance of the image stitching processes by utilizing big data framework. To realize this, we first convert the images into bitmaps (first mapper) and then String formats in the forms of 255s and 0s (second mapper), and finally, find the best possible matching position of the images by a reduce function.

## 1. INTRODUCTION

Image stitching is obtain a single image from images that have common areas with each other. Stitched images can be used in panoramic view of images, high-resolution display of mosaic images on digital maps , medical imaging and other application related to 3D environment modeling using real world images (Chia-Yen Chen, 1998), ("Image stitching - Wikipedia,",2017). Image stitching is basically divided into direct techniques and feature based techniques. Direct techniques perform operations according to pixel intensities of input images. Each pixel intensity of the image is compared with each pixel intensity of the other image. In this approach, comparing each pixel has a high complexity. In this approach, images are shifted relative to each other in order to find the degree of similarity between the pictures. These methods using pixel-pixel mapping are commonly known as direct methods (Pravenaa and Menaka, 2016). In this study, a direct technique was used in the developed algorithm. Feature based techniques solve a relationship between images based on the extracted properties of the input images (Pravenaa and Menaka, 2016), (Bonny and Uddin, 2016). For this approach; feature extraction, image registration and image blending are the following stages. There are many feature detection methods for feature based methods such as SURF, SIFT, MSER (Shaikh and Patankar, 2015), Harris, FAST (Pravenaa and Menaka, 2016). Image registration is pre-processing step which is used to merge images that at different times of the same scene.(Sayar et al., 2013). Image blending is the process of obtaining a seamless image with a smoother transition between images (Pravenaa and Menaka, 2016).

Apache Hadoop is a library of software developed with the Java programming language, which makes it possible for large datasets to be processed by clusters of computers. Hadoop is a data processing environment built on distributed file system, specially designed for very large-scale data processing (Zikopoulos, 2012). Map/reduce is a distributed programming model that consists of map and reduce steps. Users define a map function that consists of key / value pairs and a reduce function that combines all the values associated with the same key. Programs written with the map/reduce programming paradigm are automatically parallelized and can be processed on a large set (Sanjay Ghemawat, 2004).

(Eken and Sayar, 2016) aimed at creating a distributed and scalable architecture by using a bigdata framework based on map/reduce for mosaic satellite image stitcihng and object extraction. (Eken and Sayar, 2015) performed a vector-based case study to demonstrate that high performance for stitcihng satellite images can be achieved in accordance with the hadoop map reduce framework. (Sozykin and Epanchintsev, 2015) presented a distributed system by using Hadoop's map/reduce computation paradigm for image processing. Basic image processing operations such as SIFT, edge detection are distributed using Java image processing library OpenIMAJ and Java2D. In the work they do, existing libraries are made available for distributed use. (Rajak et al., 2015) presented a Hadoop map/reduce based architecture to store the program output in HBase for remote sensing satellite data. Algorithms proposed for map/reduce solution are image registration, watershed image segmentation, image mosaicing and gauss filter. Experiments on satellite data from Landsat images have shown that using Hadoop clusters to process high resolution satellite image data has a positive effect on productivity. The result is that at least 7X speed can be achieved even for complex image processing algorithms using a four-node cluster. (Vemula and Crick, 2015) have developed a library based on Hadoop map/reduce, which makes it possible to process images on a large scale. They have designed the work to abstract Hadoop's technical details of the powerful map/reduce system and provide an easy mechanism for users to manipulate large image data sets. They have developed a distributed system for Laplacian filtering, Canny edge detection and k-means image segmentation. (White et al., 2010) have implemented various practical computer vision algorithms such as classifier training, floating windows, clustering, bag-of-features, background subtraction and image registration using the map/reduce framework. (Golpayegani and Halem, 2009), (Lv et al., 2010) perform some satellite image processing algorithms using Hadoop map/reduce framework but before using images as raw in Hadoop, convert them to text format and then binary form. This pre-process has taken a lot of calculation time because they do not use the images raw. (Tesfamariam, 2011) also introduced the processing of large-sized satellite images based on map/reduce in his work and done a state study on edge detection algorithms such as Sobel, Laplacian and Canny.

Due to the positive effect on the performance of using the Hadoop map/reduce programming approach in the referenced studies, it has been decided to use Hadoop map/reduce framework for image stitching process. Apache Hadoop is a library of software developed with the Java programming language that allows large datasets to be processed by clusters of computers. Hadoop is a data processing environment on a distributed, clustered file system specially designed for very large-scale data processing ("IBM, Paul Zikopoulos, Chris Eaton, Paul Zikopoulos-Understanding Big Data Analytics for Enterprise Class Hadoop and Streaming Data-McGraw-Hill Osborne Media,",2011) .Map/reduce is a distributed programming model that consists of map and reduce steps, developed to process large sets of data on a large cluster. Users define a map function that consists of key/value pairs and a reduce function that combines all the values associated with the same key (Sanjay Ghemawat, 2004). In this study, algorithms and results are presented for creating a single new image with reference to the point where the biggest overlap index is on two pictures using Hadoop map/reduce distributed computation paradigm.

## 2. ARCHITECTURE

In this study, images are combined with the reference of the coordinate most commonly intersected with each other. The algorithm developed works on single node. The algorithm was developed in accordance with map/reduce framework using java programming language. The application architecture consists of two mapper functions running in parallel. The output of the 1.mapper function is used as an input in the 2.mapper function. While 1.map function consists of creating new space for images, 2.mapper function consists of calculating common intersections between images and two image combining processes with reference to the most common intersection coordinates calculated.

### 2.1 Conversion Image to Bitmap and String Format

Images are converted into bitmaps which represented as matrices consists of "0" and "255" elements. This image to bitmap conversion is performed by using threshold value "128" on each pixel of grayscale image. Pixel values of image are defined '0' or '255'. Black pixels are assigned '0', white pixels are assigned '255'. A bitmap matrices consisting of 0 and 255 elements is shown in Figure1.

```
255 255 255 255 255     255 255 255 255 255
255  0  255 255 255     255 255 255 255 255
255 255  0   0   0      255 255 255 255 255
255 255 255 255 255     255  0   0   0  255
255 255 255 255 255     255 255 255 255  0
```

a) Pixel Values of Image1     b) Pixel Values of Image2
Figure 1. Pixel Values of Images

In this study, bitmaps are represented as a string in order to apply input format in map function as a text. String format of image is defined as each image column seperated by a comma ',' character and each image row seperated by a semicolon ';' character. If image1 (Figure1.a) is represented as a string "255,255,255,255,255;255,0,255,255,255;255,255,0,0,0;255,255,255,255,255;255,255,255,255,255;" value is obtained.

### 2.2 Creating New Space for Images

Images are defined in the new space so that all possible matching cases between the two images can be calculated. It is the first map function to apply images to define a new space.

Input:  Image1 (String Format) <tab> Image2 (String Format)
Output: Image1 (String Format) <tab> Image2 in new Space (String Format)

```
255 255 255 255 255 255 255 255 255 255
255 255 255 255 255 255 255 255 255 255
255 255 255 255 255 255 255 255 255 255
255 255 255 255 255 255 255 255 255 255
255 255 255 255 255 255 255 255 255 255
255 255 255 255 255 255 255 255 255 255
255 255 255 255 255 255 255 255 255 255
255 255 255 255 255 255 255 255 255 255
255 255 255 255 255 255  0   0   0  255
255 255 255 255 255 255 255 255 255  0
```

Figure 2. Image2 in New Space

For each image pair, the width and height of image1 are added to the left and top of image2 as much as the white pixel. When 5x5 size images (Figure 1.a and b) are converted to the new space, the width of image1 (5 pixels) are added to the left side of image2 and height of image1 (5 pixels) are added to the top of image2. A new image with a size of 10x10 has been obtained. The new image is shown in Figure 2.

### 2.3 Calculation of the Maximum Number of Black Pixel Overlaps Between Images

With the block size created by referencing the size of the first matrix, the first matrix is traversed over the second matrix and the overlapping black pixels are counted. The elements of image1 are considered as a floating frame on image2(in new space). Black pixels in common indices are counted where one-to-one matching between the first image and the second image. The state of the floating window is shown in Figure 3. In Figure 3, the elements of the first image matrix are matched with 5x5 matrices each marked with a different color in the second image matrix. This operation is performed for all 5x5 matrices defined on the image2 and common black pixels are counted for each pairing state. The intersection coordinate, which is the highest intersection numbers obtained, is saved and used in the merging process of the next step The mathematical expression of the calculation the largest number of matches between pictures is shown by equation (1).

$$\text{Max number of intersections} = \sum_{row=0}^{n+a-1} \sum_{col=0}^{m+b-1} select\ Max$$

$$(\sum_{i=0}^{n} \sum_{i=0}^{m} 1\ [image1_{ij} = image2_{i+row,j+col} = 0]) \qquad (1)$$

The operations is performed on the 2.map function defined in the algorithm. The input of the map function is defined as the output of the 1.map function.

Input :  Image1 (String Format) <tab> Image2 in new Space (String Format)
Output : New Image Defined in String Format With The Combination of Image1 and Image2

Figure 3. Floating Frames Between 2 Images

### 2.4 Combining Images

The images are combined with reference to the largest intersection coordinate calculated between the two images. The largest intersection coordinate obtained for the first picture and the second picture merging operation is calculated as the point (4,6) on the second picture. With reference to point (4,6), the elements of image1 with a matrix size of 5x5 units are marked by the green border area shown in Figure 4.a. After considering all matching cases, the largest intersection coordinate obtained for the merging. It is calculated as the point (4,6) on the second picture. It is seen that when the matrix elements (framed by the green frame shown on image1 and image2) are overlap each other, three black pixel values are matched. The matrix elements of image1 are placed on image 2 with reference to point (4,6) on image2.

If size of image1 is assumed (nxm) and size of image2 is assumed (axb), size of merged image is defined in (n+a)x(m+b). Merging algorithm for image1 in (nxm) dimension and image2 in (axb) dimension is shown in Algorithm1.

(a) Displaying the largest intersection index

(b) New matrix obtained by merging two images.
Figure 4. Combining Images

| | |
|---|---|
| **Algorithm1:** A mapper function that computes the largest intersection between two images and performs two-image merging | |
| 1 | **class** MergeImages |
| 2 |   **method** Map(String Images) |
| 3 |     **for** row = 0 to (n + a - 1) |
| 4 |       **for** col = 0 to (m + b - 1) |
| 5 |         countMatchings ← 0 |
| 6 |         **for** i = 0 to n - 1 |
| 7 |           **for** j = 0 to m - 1 |
| 8 |             **if** image1$_{i,j}$ = image2$_{row+i, col+j}$ = 0 |
| 9 |               countMatchings ← countMatchings + 1 |
| 10 | |
| 11 |     Select max match count and coordinate(x,y) |
| 12 | |
| 13 |     **for** i = 0 to n |
| 14 |       **for** j = 0 to m |
| 15 |         image2$_{i+x,\ j+y}$ ← image1$_{i,j}$ |
| 16 |     emit(image2) |

### 2.5 Application

The developed application retrieves the file path of the images as input values from the user. After performing the reading process on the files, it converts the pictures into string format and saves them to the file. The input of the map function is the text file in which the images are saved in the string format. The map function, which performs the creating new space map function on the images and the other map function which performs the merging operation are executed one after the other. The new image is written on the file system. The program function that will convert the image of the string format to the ".png" format is executed and the combined state of the images is taken as the application output. Figure 5 shows the image inputs that taken in the application. By combining two simple images, the merge process is successful and the program output that combines the pictures is shown in Figure 6.

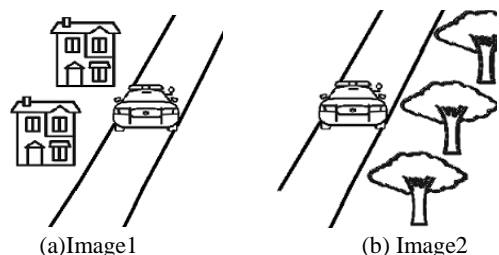

(a) Image1          (b) Image2
Figure 5. Simple images used as input data in the application

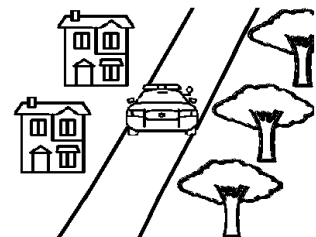

Figure 6. New image obtained by simple two-image merging.

When the merging algorithm is applied on the gray images shown in Figure 7.(a) and (b), the merging of the images has been performed (Figure 7.c).

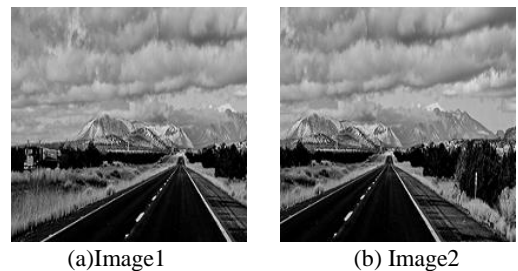

(a) Image1          (b) Image2

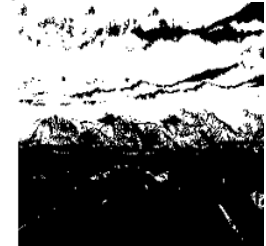

(c) Two images combined
Figure 7. Implementation of two image merging with true two gray images.

The $n^4$ complexity of the developed algorithm and its operation on a single node has not made it possible to use the performance of the algorithm in large images over time. When the time performance graph of the map functions in the application shown in Figure 8. a and b is examined, a dramatically increase in the process duration is observed with the increase of the matrix size. Although the tests made enable the processability of large images, the desired performance can not be achieved in terms of the performance of the developed application over time. The next step of the algorithm is to make it usable on large images by providing processability on a Hadoop cluster.

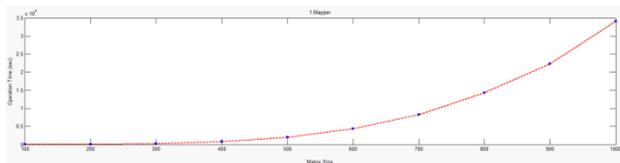

(a) Matrix size - process time (sec) graph for creating new space mapper

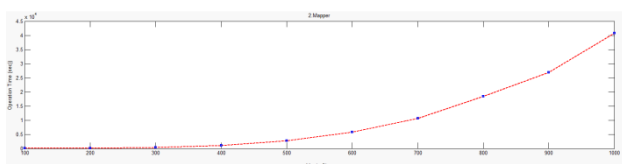

(b) Matrix size - process time (sec) graph for merging images mapper

Figure 8. Application performance graph for different matrix sizes.

## 3. CONCLUSION

In the study, an algorithm was developed to process images in accordance with Hadoop map/reduce framework. It is intended to test the performance of the map function which will work by giving an image as input to the map function expressed in text format. It is aimed to create a scalable system for large size images by reading images through the Hadoop file system and using the map/reduce programming approach. It has been tried to solve the merging process with $n^4$ complexity with reference to the point where the two images contain the most similarity with each other.

One of the improvement points of the developed algorithm is to use an algorithm that will determine the threshold value with the black-and-white pixel balance of the images as they are converted to the black-and-white pixel.The conversion of images to black and white according to the threshold value of "128" appears to be a challenge to find matches and a reduction in the accuracy of finding matches on the actual images with a high white density or black density..

The algorithm developed is based on the counting of black pixels. It is a success when the black color intensity of the pictures is low. But when the white density of the image is too high, it is difficult to catch the intersections. The next step is to add a preprocessing step that subtracts the overall color intensity of the images and the intersection will be able to positively affect the performance of the calculated application with reference to this step.

Pre-processing step of detecting the edges on the images can be effective in obtaining rapid and accurate results on the calculation of matching situations between images and the detection of the point where the images are combined. These preprocessing steps can be included in the next step.

### ACKNOWLEDGEMENTS

This work has been supported by the TUBITAK under grant 215E189.

### REFERENCES

Bonny, M.Z., Uddin, M.S., 2016. Feature-based image stitching algorithms. IEEE, pp. 198–203. doi:10.1109/IWCI.2016.7860365.

Chia-Yen Chen, 1998. Image Stitching Comparisons and New Techniques.

Eken, S., Sayar, A., 2016. A MapReduce based Big-data Framework for Object Extraction from Mosaic Satellite Images, In IoTBD 2016 International Conference on Internet of Things and Big Data (Doctoral Consortium), pp. 14-18.

Eken, S., Sayar, A., 2016. Uydu görüntülerinin yüksek performansta işlenmesi üzerine bir inceme: vektör tabanli mozaik örme durum çalişmasi, 6. Uzaktan Algılama ve CBS Sempozyumu (Turkish).

Golpayegani, N., Halem, M., 2009. Cloud Computing for Satellite Data Processing on High End Compute Clusters. IEEE, pp. 88–92. doi:10.1109/CLOUD.2009.71.

IBM, Paul Zikopoulos, Chris Eaton, Paul Zikopoulos-Understanding Big Data Analytics for Enterprise Class Hadoop and Streaming Data-McGraw-Hill Osborne Media ,2011.

Image stitching - Wikipedia [WWW Document], URL https://en.wikipedia.org/wiki/Image_stitching (accessed 8.1.17).

Lv, Z., Hu, Y., Zhong, H., Wu, J., Li, B., Zhao, H., 2010. Parallel K-Means Clustering of Remote Sensing Images Based on MapReduce, in: Wang, F.L., Gong, Z., Luo, X., Lei, J. (Eds.), Web Information Systems and Mining. Springer Berlin Heidelberg, Berlin, Heidelberg, pp. 162–170. doi:10.1007/978-3-642-16515-3_21.

Pravenaa, S., Menaka, R., 2016. A Methodical Review on Image Stitching and Video Stitching Techniques. Int. J. Appl. Eng. Res. 11, 3442–3448.

Rajak, R., Raveendran, D., Bh, M.C., Medasani, S.S., 2015. High Resolution Satellite Image Processing Using Hadoop Framework. IEEE, pp. 16–21. doi:10.1109/CCEM.2015.16.

Sanjay Ghemawat, J.D., 2004. MapReduce:Simplified Data Processing on Large Clusters. Presented at the Proceedings of the 6th conference on Symposium on Opearting Systems Design and Implementation - Volume 6.

Sayar, A., Eken, S., Mert, U., 2013. Registering landsat-8 mosaic images: A case study on the Marmara Sea, in: Electronics, Computer and Computation (ICECCO), 2013 International Conference on. IEEE, pp. 375–377.

Shaikh, T.S., Patankar, A.B., 2015. Multiple Feature Extraction Techniques in Image Stitching. Int. J. Comput. Appl. 123.

Sozykin, A., Epanchintsev, T., 2015. MIPr - a framework for distributed image processing using Hadoop, in: 2015 9th

International Conference on Application of Information and Communication Technologies (AICT). Presented at the 2015 9th International Conference on Application of Information and Communication Technologies (AICT), pp. 35–39. doi:10.1109/ICAICT.2015.7338511.

Tesfamariam, E.B., 2011. Distributed processing of large remote sensing images using MapReduce-A case of Edge Detection.

Vemula, S., Crick, C., 2015. Hadoop Image Processing Framework, in: 2015 IEEE International Congress on Big Data. Presented at the 2015 IEEE International Congress on Big Data, pp. 506–513. doi:10.1109/BigDataCongress.2015.80.

White, B., Yeh, T., Lin, J., Davis, L., 2010. Web-scale computer vision using mapreduce for multimedia data mining, in: Proceedings of the Tenth International Workshop on Multimedia Data Mining. ACM, p. 9.